\documentstyle[prb,aps,epsf]{revtex}
\draft

\begin{document}

%%%  TWO-COLUMN FORMAT and EPSF %%%%%%%
\twocolumn[\hsize\textwidth\columnwidth\hsize
    \csname @twocolumnfalse\endcsname
%%%  TWO-COLUMN FORMAT and EPSF %%%%%%%
\vspace{-1.7cm}
%%%\rule{10.3cm}{0cm} {\Large\bf{PRELIMINARY DRAFT}}\\
%\rule{9.7cm}{0cm} {\Large\bf{Phys. Rev. B (to be submitted)}}
\rule{9.7cm}{0cm} {\Large\bf{Phys. Rev. B (submitted, 2001)}}
%\rule{8.5cm}{0cm} {\Large\bf{Phys. Rev. Lett. (submitted, 2001)}}
\vspace{+0.7cm}
%%%\vspace{+0.3cm}
%%%%%%%%%%%%%%%%%%%%%%%%%%%%%%%%%%%%%%%%%%%%%%%%%%%%%%%%%%%%%%%%%%%%%%%%%%%%%%%%
 
%%%%%%%%%%%%%%%%%%%%%%%%%%%%%%%%%%%%%%%%%%%%%%%%%%%%%%%%%%%%%%%%%%%%%%
\title{Fully relativistic calculation of magnetic properties of
Fe, Co and Ni adclusters on Ag(100)}
\author{B. Lazarovits$^{1}$, L. Szunyogh$^{1,2}$ and
P. Weinberger$^1$}
\address{$^1$Center for Computational Materials Science,
Technical University Vienna,
\\ A-1060 Gumpendorferstr. 1.a., Wien, Austria }
\address{$^2$ Department of Theoretical Physics,
Budapest University of Technology, \\
Budafoki \'ut 8, H-1521, Budapest, Hungary}
%%%%%%%%%%%%%%%%%%%%%%%%%%%%%%%%%%%%%%%%%%%%%%%%%%%%%%%%%%%%%%%%%%%%%%

%%%  TWO-COLUMN FORMAT and EPSF %%%%%%%
%\date{submitted to Phys. Rev. B}
%%%  TWO-COLUMN FORMAT and EPSF %%%%%%%

\maketitle

%%%%%%%%%%%%%%%
%   ABSTRACT  %
%%%%%%%%%%%%%%%

\begin{abstract}

We present {\em first principles} calculations of
the magnetic moments and magnetic anisotropy energies
 of small Fe, Co and Ni clusters on top of a Ag(100) surface
 as well as the exchange-coupling energy between two single adatoms of
Fe or Co  on Ag(100).
The calculations are performed
fully relativistically using the
 embedding technique within the Korringa--Kohn--Rostoker method.
The magnetic anisotropy and the exchange--coupling energies are
calculated by means of the force theorem.
In the case of adatoms and dimers of iron and cobalt
we obtain enhanced spin moments and, especially,
unusually large orbital moments, while for
nickel our calculations predict a complete absence of magnetism.
For larger clusters, the magnitudes of the local moments
of the atoms in the center of the cluster
are very close to those calculated for the corresponding monolayers.
Similar to the orbital moments, the contributions of the
individual atoms to the magnetic anisotropy energy
strongly depend on the position, hence, on the local environment
of a particular atom within a given cluster.
We find strong ferromagnetic coupling between two neighboring Fe or Co atoms
and a rapid, oscillatory decay of the exchange-coupling energy with
increasing distance between these two adatoms.

\end{abstract}

%%%  TWO-COLUMN FORMAT and EPSF %%%%%%%
\vskip2pc]
%%%  TWO-COLUMN FORMAT and EPSF %%%%%%%

\narrowtext

%%%%%%%%%%%%%%%%
% INTRODUCTION %
%%%%%%%%%%%%%%%%
\section{Introduction}
\label{sec:intro}

Magnetic nanostructures such as impurities,
clusters and wires on top
 or in the uppermost layers of surfaces are of special interest
 for nano-scale technologies, in particular,
 regarding their possible application as magnetic nano-devices and
high-density magnetic recording media.
A quantitatively correct description of the magnetic properties
of such structures, namely, the
magnitude and the orientation of spin and orbital moments, magnetic anisotropy
energies and the magnetic interactions,
 is, therefore, an important issue to be addressed.
Concomitantly, the understanding of the changes of physical properties
from nanostructures to thin films or even bulk systems 
has always been a fascinating theoretical challenge.

Because of the lack of translational symmetry
tight-binding (TB) methods have been an efficient tool to study
larger clusters. By using a tight--binding
Hubbard Hamiltonian in the unrestricted Hartree--Fock approximation,
Pastor and co--workers revealed the size and structural dependence
of magnetic properties of free Cr$_n$, Fe$_n$ and Ni$_n$ ($n \le 15$)
clusters \cite{PDB89}, and also the exchange interaction and local environments 
effects in Fe$_n$ clusters \cite{DDP92}. By including a spin--orbit
coupling term into the Hamiltonian, they also investigated various
effects on the magnetic anisotropy energy (MAE) of small unsupported 
Fe clusters \cite{PDP+95} and, recently, of Co$_n$ clusters
on Pd(111) \cite{FGD+00}. Finite temperature magnetism of
small clusters, remarkably different from that of bulk systems,
has also been studied in terms of a similar approach by taking into
account both electronic and structural excitations \cite{LPB00}.
A great advantage of the TB methods seems to be that they easily can be combined
with molecular dynamics calculations enabling thus investigations
of relaxation effects which proved to be important in determining the
magnetic moments \cite{PDO+96a,RAV+99,GDS00} and the MAE \cite{G01} of
transition metal clusters.

The embedding technique based on the 
Korringa--Kohn--Rostoker (KKR) Green's function method 
in the local spin-density approximation (LSDA)
has been applied 
to the magnetism of transition metal adatoms and clusters deposited on surfaces  
\cite{LSW+94,WSL+95}. The main feature of this approach
is that the interaction between adatoms and host surface
atoms can be analyzed 
within {\em first principles} electronic
structure calculations \cite{SHW+96,SHR+96}, in several cases exhibiting
novel phenomena in nanomagnetism such as the existence of metamagnetic
states \cite{SHR+97a,SHR+97b} or intermixing effects between
adatoms and the host surface \cite{SHR+99}.
An accurate calculation of the total energy in terms of full potential
or full charge density schemes made possible the investigations
of the energetics of adatoms \cite{NWZ+98,LSH+00,SH00}. 
As compared to TB methods an obvious drawback of the embedded
KKR technique is that, with respect to computational limitations,
the number of the atoms in the cluster is  
restricted to about less than 100.  
Furthermore, the inclusion of structural relaxations is exceedingly
difficult. In order to circumvent these problems, a quasi-ab
initio molecular--dynamics method has been employed by 
parameterizing interatomic potentials to the first principles
KKR Green's function electronic structures \cite{IBV+01}.
On the level of a fully relativistic spin-polarized electron theory,
recently, strongly enhanced orbital magnetism and MAE
of adatoms and small clusters on Ag and Au(100) surfaces have
been reported \cite{NCZ+01,CNZ+01}.

From the mid-nineties on we carried out systematic investigations 
of the magnetism, in particular of the MAE, of transition
metal multilayer systems by using the fully relativistic spin-polarized 
screened Korringa--Kohn--Rostoker 
(SKKR) method \cite{SUW+94a,ZDU+95,WS00}.
Specifically, within the single-site approximation,
we explored the oscillatory behavior of the MAE of 
an Fe impurity buried in a Au(100) host \cite{SG97a}.
The purpose of the present work is to extend these studies by including
self--consistent effects (electronic relaxations) of the host atoms
in order to perform realistic investigations for magnetic
clusters on metallic surfaces. For this very reason we make use of
a real--space embedding technique in order to calculate the electronic structure 
of the cluster, and also to be able to treat the Poisson equation with appropriate
boundary condition. Theoretical and computational details 
are given in Sections \ref{sec:method} and \ref{sec:comdet},
respectively. In Section \ref{sec:results} 
our results of the magnetic moments and the MAE of small planar
Fe, Co and Ni clusters on Ag(100), as well as of the magnetic
correlation between  Fe or Co adatoms are presented.
Finally, in Section \ref{sec:summary} we summarize and draw conclusions.

%%%%%%%%%%%%%%%%%%%%%%%%%%%%%%%%%%%%
% THEORY AND COMPUTATIONAL DETAILS %
%%%%%%%%%%%%%%%%%%%%%%%%%%%%%%%%%%%%
%\section{Theory and computational details}
\section{Theoretical approach}
\label{sec:method}

Within multiple scattering theory the scattering path operator 
(SPO) matrix, $\mbox{\boldmath $\tau$}(E)=\{\underline{\tau}^{nm}(E)\}=
\{\tau_{QQ'}^{nm}(E)\}$, with $Q$ and $Q'$ being 
angular momentum indices and $E$ being the energy, defined as
\begin{equation}
     \mbox{\boldmath $\tau$}(E)=
           [
            \mbox{\boldmath $t$}^{-1}(E)-
            \mbox{\boldmath $G$}(E)
           ]^{-1} 
\quad,
\label{eq:realtau}
\end{equation} 
describes the full hierarchy of scattering effects between any two
particular sites, $n$ and $m$. In Eq.~(\ref{eq:realtau}),
$\mbox{\boldmath $t$}(E)=\{\underline{t}^{n}(E)\, \delta_{nm}\}=
\{t_{QQ'}^{n}(E) \, \delta_{nm}\}$ and
$\mbox{\boldmath $G$}(E)=\{\underline{G}^{nm}(E)\}= \{G_{QQ'}^{nm}(E)\}$ 
denote the single--site {\em t}-matrices and the real-space 
structure constants, respectively.
For more details, especially, how to calculate
$t_{QQ'}^{n}(E)$ within a fully relativistic 
spin--polarized scheme, see, e.g., Ref. \cite{weinbook}.

Assuming that a finite set of impurities interacts within 
a given finite range, we can select an environment of  
impurities, $\cal{C}$, containing also perturbed
host atoms, such that outside $\cal{C}$ the potentials 
can be considered to be identical with those of the unperturbed host.
A particular cluster $\cal{C}$ can then be treated as 
perturbation of the host.
In practice, we first calculate the SPO of the 2D translational 
invariant layered host, 
${\mbox{\boldmath $\tau$}}_{h}({\bf k}_{\parallel},E)=
\{\underline{\tau}_{h}^{pq}({\bf k}_{\parallel},E)\}$, within the 
framework of the SKKR method\cite{SUW+94a}, where $p$ and $q$ denote
layers and the ${\bf k}_{\parallel}$ are vectors in the surface
Brillouin zone (SBZ).
The real--space SPO is then given by 
\begin{equation}
    \underline{\tau}^{mn}_{h}(E)=
    \frac{1}{\Omega_{SBZ}}
    \int\limits_{SBZ}e^{-i({\bf T}_i-{\bf T}_j)
    {\bf k}_{\parallel}}
    \underline{\tau}^{pq}_{h}(
    {\bf k}_{\parallel} ,E)
    d^2k_{\parallel}
\label{eq:bzint}
\quad,
\end{equation}
where the atomic position vectors are decomposed as 
${\bf R}_m={\bf T}_i+{\bf c}_p$ and 
${\bf R}_n={\bf T}_j+{\bf c}_q$ with ${\bf T}_i$
and ${\bf T}_j$ being 2D lattice vectors, 
${\bf c}_p$ and ${\bf c}_q$ the so--called layer--generating
vectors, and $\Omega_{SBZ}$ is the unit area of the surface Brillouin zone. 

By replacing the {\em t}--matrices of the unperturbed host,
$\mbox{\boldmath $t$}_{h}(E)$, with those of the cluster-atoms,
$\mbox{\boldmath $t$}_{\cal{C}}(E)$, 
leads to the following Dyson like equation,
\begin{equation} 
    \mbox{\boldmath $\tau$}_{\cal{C}}(E)=
    \mbox{\boldmath $\tau$}_{h}(E)
       \left[
          \mbox{\boldmath $I$}-
          (\mbox{\boldmath $t$}^{-1}_{h}(E) - 
           \mbox{\boldmath $t$}^{-1}_{\cal{C}}(E))
         \mbox{\boldmath $\tau$}_{h}(E)
        \right]^{-1}
\quad ,
\label{eq:dyson}
\end{equation}
where $\mbox{\boldmath $\tau$}_{\cal{C}}(E)$ is the 
SPO-matrix corresponding to all sites in  
cluster ${\cal C}$, from which in turn all corresponding local
quantities, i.e., charge and magnetization densities, spin-- and
orbital moments, as well as the total energy can be calculated. 
Note, that Eq.~(\ref{eq:dyson}) takes into account all 
scattering events both inside and outside the cluster. 

As usual, in order to perform self-consistent calculations, the Poisson equation has to be 
solved repeatedly. Partitioning the configurational space into non-overlapping cells
centered at the atomic sites only the {\em intracell}
part needs special care. 
By using a multipole expansion of the Coulomb potential, one arrives at 
 the so--called Madelung potentials, which for a given 
site $n$ of ${\cal{C}}$ will be denoted by $V^{M}_n$.
Due to the additivity of the Poisson equation, $V^{M}_n$ 
can be decomposed into contributions from 
atoms inside and outside of the cluster,
$V^{M}_{{\cal{C}},n}$ and $V^{M}_{{\cal{S}},n}$, 
respectively,
\begin{equation}
     V^{M}_n= V^{M}_{{\cal{C}},n}+
              V^{M}_{{\cal{S}},n} \; ,
\label{eq:clupot:sep}
\end{equation}
where ${\cal{S}}$ denotes the ensemble of scatterers 
outside ${\cal{C}}$.
Clearly enough, assuming that the ensemble of sites in  
$\cal{S}$ is independent of $\cal{C}$, 
implies that $V^{M}_{{\cal{S}},n}$ has to be independent of
the type of the atoms in $\cal{C}$. 
Replacing the atoms in $\cal{C}$
by those of the unperturbed host, one can write
\begin{equation}
     V^{M}_{{\cal{S}},n}=
     V^{M,h}_n-
     V^{M,h}_{{\cal{C}},n} \; ,
\label{eq:clupot:out}
\end{equation}
where $V^{M,h}_n$ denotes the Madelung potential of site $n$ 
for the unperturbed,
2D translational invariant host \cite{SUW+94a} and
$V^{M,h}_{{\cal{C}},n}$ stands for the Madelung potential
of site $n$ generated by
cluster ${\cal{C}}$ occupied by unperturbed host atoms.
Substituting Eq.~(\ref{eq:clupot:out}) into (\ref{eq:clupot:sep}), yields
\begin{equation}
     V^{M}_n= V^{M}_{{\cal{C}},n} - 
              V^{M,h}_{{\cal{C}},n} +
              V^{M,h}_n \quad \; ,
\label{eq:clupot}
\end{equation}
which can be regarded as an embedding equation for the intercell
potential resulting from the boundary condition set by the size of \(\cal{C}\).
Obviously, by using Eq.~(\ref{eq:clupot}) the problem of summing up
the contributions to the
intercell potential from region ${\cal{S}}$ is properly solved.

\section{Computational details}
\label{sec:comdet} 
Self--consistent, fully relativistic calculations for selected, planar Fe, Co and Ni
clusters on Ag(100) have been performed in the framework of the local
spin--density approximation as parameterized by Vosko {\em et al.} \cite{VWN80}. 
In each case three different orientations for the magnetization were 
considered:
along the $z$ axis (normal to planes), as well as along the $x$ and $y$
axes (nearest neighbor directions in an fcc(100) plane).
The potentials were treated within the atomic sphere approximation (ASA).
For the calculation of the $t$--matrices and for the multipole 
 expansion of the charge densities, necessary to evaluate the
Madelung potentials, a cut--off of $\ell_{max}=2$ was used.
In order to perform the energy integrations, 16 points on a semicircular contour 
in the complex energy plane were sampled according to an asymmetric Gaussian
quadrature. Both, for the self--consistent calculation of the Ag(100)
surface and for the evaluation of Eq.~(\ref{eq:bzint}) we used 
45 $k_{\parallel}$--points in the irreducible wedge of the SBZ.
For some restricted cases we checked the convergence of the results by
increasing the number of $k_{\parallel}$--points to 210. 

In the present study we made no attempts to include lattice 
relaxation effects, therefore, the host and the cluster sites
refer to positions of an ideal fcc parent lattice~\cite{W97} with the
experimental Ag lattice constant (4.12 \AA). 
Three layers of self--consistently treated empty sites were used
to represent the vacuum region \cite{SUW+94a}; 
the magnetic adatoms occupy sites in the first
vacuum layer. As shown in Fig.~\ref{fig:clusters}, 
we considered dimers and linear trimers oriented along
the $x$ axis, square--like tetramers,    
centered pentamers (as in Ref.~\cite{CNZ+01}), 
as well as a cluster arranged on positions
of a 3$\times$3 square denoted in the following simply 
as 3$\times$3 cluster. 
In Fig.~\ref{fig:clusters}, for each particular cluster
the equivalent atoms with respect to an orientation of the 
magnetization along the $x$ or $y$ axis
are labelled by the same number. Note that for a magnetization
aligned in the $z$ direction, the atoms labelled by 2 and 3 
in the pentamer and the 3$\times$3 cluster become equivalent.
Up to a total of 67 sites, the clusters consisted of 
adatoms, several substrate Ag atoms and empty sites from neighboring shells.
 A stability test of the local electronic and magnetic
properties for a single Fe adatom with
respect to the number of self--consistently treated neighboring
shells is shown in Fig.~\ref{fig:shells}. 
Although the calculated orbital moment of the Fe adatom shows a somewhat 
slower convergence than the valence charge and the spin moment, 
it is remarkable that considering only a first
shell of neighbors this already yields values which differ by less than 1 \% 
from the fully converged ones. 

%%%%%%%%%%%%%%%%%%%%%%%%%%%%
%  RESULTS AND DISCUSSION  %
%%%%%%%%%%%%%%%%%%%%%%%%%%%%
%       Local moments      %
%%%%%%%%%%%%%%%%%%%%%%%%%%%%
\section{Results and discussion}
\label{sec:results}

\subsection{Spin and orbital moments}
\label{sec:moments}

Calculations for different orientations of the 
magnetization revealed that 
the spin moments are fairly insensitive to the direction
of the magnetization, while for the orbital moments 
remarkably large anisotropy effects apply, a phenomenon that will
be discussed in the next Section. 
For a magnetization along the $z$ axis, the calculated values of the 
spin and orbital moments for an adatom and selected clusters of 
Fe, Co and Ni on Ag(100)
are listed in Table~\ref{tab:moments}. In there the position indices
in a particular cluster refer to the corresponding numbers in
Fig.~\ref{fig:clusters}
and the number of nearest neighbors 
of magnetic atoms (coordination number, $n_c$) is also given.

As compared to the corresponding monolayer values (3.15~\(\mu_B\) for Fe and
2.03~\(\mu_B\) for Co), the 
spin moment of a single adatom of Fe (3.39~\(\mu_B\)) 
and Co (2.10~\(\mu_B\)) is slightly increased.
In the case of Fe clusters,
the spin moments decrease monotonously with increasing $n_c$. 
A slight deviation from that behavior can be seen for the 3$\times$3 cluster,
where the atoms with $n_c$=2 and 3 exhibit the same spin moment.
For the central atom of the pentamer and, in particular, of the
3$\times$3 cluster, the monolayer value is practically achieved.
The above results compare fairly well to those of Cabria {\em et al.}
\cite{CNZ+01} and reflect a very short ranged magnetic correlation 
between the Fe atoms. 

The general tendency of decreasing spin moments with increasing
$n_c$ is obvious also for the Co clusters up to the pentamer case. 
For the 3$\times$3 cluster, however, just the opposite trend applies.
Establishing a correlation between $S_z$ and $n_c$ for
Co seems to be more ambiguous than for Fe, because the changes of the
spin moment are much smaller in this case.
Nevertheless, it is tempting to say that 
 in the formation of the magnetic moment of Co, further off
neighbors play a more significant role than in the case of Fe.

In the case of an adatom and dimer of Ni we
found no stable magnetic state. Quite contradictory,     
Cabria {\em et al.} \cite{CNZ+01} reported a spin moment of about 
0.5~$\mu_B$ for a Ni adatom on Ag(100). 
As the computational method of these authors
is very similar to ours, it is at present puzzling what causes this 
discrepancy between the two calculations.
Our result is, however, in line with the experiments 
of Beckmann and Bergmann who found
no magnetic moment for Ni adatoms on Au surface\cite{BB96a},
which as a substrate is rather similar to Ag.
It should be noted, however, that in Ref.~\cite{BB96a} the actual surface
orientation is not specified. 

For clusters of Ni one can observe an opposite tendency as for
Fe and Co: the spin moment enhances with increasing number of neighbors. 
This clearly can be seen from Table \ref{tab:moments}. 
Having in mind
the calculated monolayer value ($0.71~\mu_B$), our small cluster
calculations indicate  a fairly slow  
evolution of the spin moment of Ni with increasing cluster size, 
implying that the magnetism of Ni is subject to
correlation effects on a much longer scale than in Fe or Co.

Apparently, the orbital moments show a different, in fact,
more complex behavior as the spin moments.
For single adatoms of Fe and Co we found orbital moments 
enhanced by a factor of $\sim6$ and $\sim4.5$, respectively, 
as compared to the monolayer values ($0.14~\mu_B$ for Fe 
and $0.27~\mu_B$ for Co).
This is a direct consequence of the reduced crystal field splitting,
being relatively large in monolayers, and, in particular, in corresponding bulk
systems \cite{Bruno93}.
In spite of a qualitative agreement, our $L_z$ values for the adatoms
are considerably larger than those calculated by Cabria {\em et al.}
\cite{CNZ+01} ($0.55 \mu_B$ for Fe and $0.76 \mu_B$ for Co). 
It should be noted, however, that by including orbital polarization effects
(Hund's second rule) in terms of Brooks' parameterization
\cite{B85,EB96}, Nonas {\em et al.} \cite{NCZ+01}
found orbital moments for Fe and Co adatoms on Ag(100)
close to the atomic limit ($2.20 \mu_B$ for Fe and $2.57 \mu_B$ for Co).

For dimers of Fe and Co, the value of $L_z$  drops to about 40~\%
in magnitude as compared to a single adatom. The evolution of
the orbital moment seems, however, to decrease explicitly only for the central atom
of larger clusters. 
In a previous paper~\cite{SUW97} we showed that the (local)
symmetry can be correlated with the magnetic anisotropy, 
i.e., with the quenching effect of the crystal field 
experienced by an atom. 
A single adatom and the central atom of the 
linear trimers, pentamers and the 3$\times$3 clusters
exhibit well--defined rotational symmetry,
namely, $C_1$, $C_2$, $C_4$ and $C_4$, respectively. The corresponding
values of $L_z$, namely, 0.88 $\mu_B$, 0.25 $\mu_B$, 0.15 $\mu_B$,
and 0.12 $\mu_B$ for Fe, and 1.19 $\mu_B$, 0.49 $\mu_B$, 0.25 $\mu_B$
and 0.23 $\mu_B$ for Co, nicely reflect the
increasing rotational symmetry of the respective atoms.
Although the outer magnetic atoms exhibit systematically
larger orbital moments than the central ones, even a qualitative
correlation with the local environment ($n_c$) can hardly be stated. 
The orbital moment for the trimer of Ni is already close to the monolayer value 
(\(0.19~\mu_B\)) but shows rather
big fluctuations with respect to the size of the cluster and
also to the positions of the individual atoms.  

%%%%%%%%%%%%%%%%%%%%%%
% Anisotropy energy  %
%%%%%%%%%%%%%%%%%%%%%%
\subsection{Anisotropies of orbital moments and magnetic anisotropy
energies}
\label{sec:anis}

By using the self--consistent potentials for a given orientation
of the magnetization (along $z$), we calculated 
magnetic anisotropy energies by means of
the magnetic force theorem \cite{SUW95,Jan99} 
as differences of band--energies,
\begin{equation} \label{eq:MAE}
\Delta E_{x-z}=E_{b;x}-E_{b;z} \quad {\rm and} \quad
\Delta E_{y-z}=E_{b;y}-E_{b;z} \; .
\end{equation}
For a particular orientation $\alpha$, the band--energy 
is obtained as a sum of contributions from all atoms in the cluster,
\begin{equation} \label{eq:bande-clus}
E_{b;\alpha} = \sum_{i \in {\cal C}} E_{b;\alpha}^i \quad 
(\alpha=x,y,z) \; ,
\end{equation}
\begin{equation} \label{eq:bande-atom}
E_{b;\alpha}^i = \int^{\epsilon_F}_{\epsilon_B} \! d\epsilon 
\: (\epsilon - \epsilon_F) \: n_{\alpha}^i(\epsilon) \; ,
\end{equation}
where $\epsilon_F$ is the Fermi energy of the substrate, 
$\epsilon_B$ is the bottom of the valence band and
$n_{\alpha}^i(\epsilon)$ is the density of states for atom 
 $i$. Clearly, the above formalism allows us to define
the MAE as a sum of atom--like contributions, which facilitates
to trace its spatial distribution in the cluster. 

The anisotropies of the orbital moments and the contributions of
the individual magnetic atoms to the MAE are displayed in
Tables~\ref{tab:feani}, \ref{tab:coani} and 
\ref{tab:niani} for Fe, Co and Ni clusters, respectively. 
%Note that an in--plane ($x$ or $y$) orientation of
%the magnetization reduces the rotational symmetry of the clusters
%as compared to an orientation along the $z$ axis. Therefore, particular sites
%in the trimers, pentamers and the 3$\times$3 clusters, that are 
%equivalent in the case when
%the magnetization is oriented along \(z\), can be further resolved. 
In addition, the total MAE per magnetic atoms of the clusters including the
neighborhood is also given.
Although the dominating contributions to the MAE
arise from the magnetic species, the environment, in particular,
the Ag atoms and the empty sites within the first shell add a 
remarkable amount to the MAE. 
However, due to the weak polarization of the Ag atoms, 
we obtained a fast convergence of the total MAE
with respect to the size of the cluster (environment).

As can be inferred from the corresponding positive values of
the MAE in Tables~\ref{tab:feani} and \ref{tab:coani},
single adatoms of Fe and Co exhibit a magnetization oriented perpendicular
to the surface.
This again is in perfect agreement with the experiments of Beckmann
and Bergmann~\cite{BB96a}. As compared with the monolayer case
(0.47 meV), the MAE of an Fe adatom (5.61 meV) is enhanced by a factor of 
twelve. Contrary to our results, 
Cabria {\em et al.} \cite{CNZ+01} predicted in--plane magnetism
($\Delta E_{x-z}$ = -0.98 meV) for an Fe adatom on Ag(100),
and perpendicular magnetism for Co, albeit with a much larger
anisotropy energy ($>$ 7 meV) than ours (4.36 meV). 
It should be stressed at this point, that our calculations are consistent
with a qualitative rule, valid for transition metals with a more than 
half--filled $d$--band and based on simple, perturbative phenomenological or
tight--binding reasoning \cite{Bruno93}: the direction,
along which the orbital moment is the largest, is energetically favored.

As can be seen from Table~\ref{tab:feani}, perpendicular magnetism is 
characteristic for all Fe clusters considered. For the dimer and
the trimer we observe a small in--plane anisotropy with preference
of the $x$ axis, i.e., in the direction of the Fe-Fe bonds. 
In agreement with the reduction of the orbital moment, as discussed
in the previous section, the contribution of the central atom to the MAE 
for the trimer, the pentamer and the 3$\times$3 cluster rapidly decreases, 
being even less than the monolayer value in the case of the 3$\times$3
cluster.  The outer atoms in the pentamer and in the 3$\times$3 cluster
 can add considerably more
to the MAE than the central atom. As a consequence, the average MAE
strongly fluctuates with increasing size of the magnetic cluster
and shows a very slow tendency to converge to the
MAE of an Fe monolayer on Ag(100). Such a complicated behavior of
the MAE with respect to the cluster size has also been found 
by Guirado--L\'opez \cite{G01} for free--standing fcc
transition metal clusters.

In comparison to an adatom, for a Co dimer $\Delta E_{x-z}$ 
drops to a large negative value (-3.49 meV/per Co), while $\Delta E_{y-z}$ 
remains slightly positive (0.76 meV/per Co), implying that a
Co dimer favors the $x$ (in--plane) direction of the magnetization 
and also experiences a strong in--plane anisotropy. The strong  
tendency of Co clusters to in--plane magnetization pertains to
larger clusters and is characteristic also for a Co monolayer
($\Delta E_{x-z}$ = -1.31 meV). The atom--like
resolution of the MAE indicates, that this tendency is driven
by nearest--neighbor Co-Co interactions.
An explanation of this
effect in terms of perturbation theory and symmetry resolved
densities of states can be found in Refs.~\cite{WWF94,ZKW+96,USB+96}.
As an unexpected consequence, 
the contribution to the MAE of the central atom in the cluster 
can be larger than that of some outer atoms.
Quite obviously, the MAE of the central atom of the trimer, 
the pentamer and the 3$\times$3 cluster, 
-9.06 meV, -2.46 meV, and -1.86 meV, respectively,
fall monotonously off to the monolayer value, whereas the average MAE
possesses a much more complicated evolution also in this case. 

With exception of the tetramer, for which we found a MAE close to zero,
all Ni clusters prefer an in--plane magnetization. The in--plane
anisotropy, seen from Table~\ref{tab:niani} for the trimer, but also 
from the atom--like contributions for the larger
clusters, is, however, smaller than in the case of Co. 
Again the complicated nature of the magnetism of Ni shows up,
in particular, for the 3$\times$3 cluster: while the contribution of the
central atom to the MAE almost vanishes, those of the outer atoms
oscillate in magnitude. Considering the MAE of a Ni monolayer on
Ag(100) (-2.23 meV), no straightforward connection with the
magnetic anisotropy properties of small clusters can be traced.

%%%%%%%%%%%%%%%%%%%%%%%%%%%%%
% Exchange-coupling energy  %
%%%%%%%%%%%%%%%%%%%%%%%%%%%%%
\subsection{Magnetic interaction between adatoms}
\label{sec:inter}

Interactions between magnetic nanoclusters are of great importance for 
technological applications. Clearly enough the most important
questions are $(i)$ what is the magnetic structure of
the individual entities,
$(ii)$ of what nature (strength, range, etc.) is the coupling 
between them, and $(iii)$  what influences the magnetic orientation 
of these entities relative to each other.
In this section we present a preliminary study in this field 
by investigating the interaction of two Fe or Co adatoms on Ag(100). 

We first performed self--consistent calculations for two adatoms
by varying the distance $d$ between them from $a$ to $5a$
along the $x$ direction, where $a$ is the 2D lattice constant and
keeping the orientation of the magnetizations
parallel to each other (along the $z$ axis).
The calculated spin and orbital moments of the (coupled) adatoms
are shown in Fig.~\ref{fig:moments2feco}. Note that the distance
$a$ refers to the bondlength in dimers. As can be seen from 
Fig.~\ref{fig:moments2feco}, both for Fe and Co the values of
$S_z$ and $L_z$ rapidly converge to the respective
single adatom value. 

Next we calculated the exchange-coupling energy, $\Delta E_X$,
between the two adatoms 
by taking the energy difference between a parallel 
($\uparrow \uparrow$) and an antiparallel ($\uparrow \downarrow$)
orientation of the two adatoms,
\begin{equation}
\Delta E_{X}=E_{b}(\uparrow\uparrow)-E_{b}(\uparrow\downarrow) \; .
\end{equation}
using, however, the self--consistent potentials for the 
parallel configuration.  We are aware of the fact that, 
due to the lack of self--consistency in the antiparallel case, 
for near adatoms this approach might be quite poor. 
We believe, however, that this approximation provides
a good estimate of the sign and the magnitude of the interaction.

The calculated $\Delta E_X$ is shown in Fig.~\ref{fig:xc} 
for Fe and Co
 as a function of the distance $d$ between the two adatoms. 
Apparently, for $d=a$ in both cases a strong, 
ferromagnetic nearest--neighbor
exchange-coupling between these two atoms applies,
with an interaction energy somewhat larger for Fe than for Co.
As the two adatoms are adjacent in this case, this strong coupling 
can be attributed to a direct exchange mechanism.
Increasing the separation between the two adatoms, $\Delta E_X$ rapidly
decreases.  For $d=2 a$ it changes sign, i.e., the coupling
becomes antiferromagnetic.
Since for an antiparallel alignment of the
spin moments of the two adatoms, lying close to each other,
the electronic structure and the magnetic moments might be
expected to differ to some extent as compared 
to a parallel configuration, the corresponding
values of $S_z$ and $L_z$ in Table~\ref{fig:moments2feco} can be
questioned. Therefore, for this particular case we performed
self--consistent calculations also for the antiparallel alignment.
Assuringly, for both Fe and Co, 
we obtained the same value of $S_z$ and $L_z$
within 1\% relative accuracy 
as in the case of a parallel alignment.
For larger distances we observe
ferromagnetic coupling, which virtually vanishes for $d \ge 5a$,
implying a very weak, short ranged exchange interaction between
the adatoms of Fe and Co induced by the Ag host.

%%%%%%%%%%%
% Summary %
%%%%%%%%%%%

\section{Summary and conclusions}
\label{sec:summary}

By using a real--space embedding technique based on the 
Korringa--Kohn--Rostoker Green's function method, we have performed 
fully relativistic, self-consistent calculations for 
adatoms and small clusters of Fe, Co and Ni on  
Ag(100). Due to the decreased coordination of the magnetic
atoms, we obtained slightly enhanced spin moments for adatoms and
small clusters of Fe and Co and
found that the spin moments are already close to the monolayer values
for a cluster of 9 atoms.
In agreement with experiments~\cite{BB96a}
the adatoms and dimers of Ni turned out to be nonmagnetic, while
the spin moments in larger Ni clusters indicated
a complex formation of magnetism.
In connection with strongly enhanced orbital moments, for Fe and Co
adatoms we revealed an unusually strong tendency to perpendicular
magnetism.
The perpendicular magnetism persisted also for Fe clusters of 
increasing size, whereby the atom--like contributions showed
an oscillating behavior depending mainly on the local rotational symmetry.
The preferred orientation for clusters of Co and 
Ni obtained was in--plane. 
In addition, we investigated the magnetic coupling between two
adatoms of Fe or Co, for which we established a 'good local--moment' 
behavior. In terms of calculated exchange-coupling energies, the dimers 
show a strong ferromagnetic coupling, which immediately drops
two orders of  magnitude with increasing distance between the two adatoms,   
indicating a weak, indirect coupling between them.

The main outcome of the present paper is
that by performing {\em first principles} calculations,
not only the qualitative trends of small cluster magnetism of
transition metals, but even quantitative results can be obtained
which in turn can directly be compared with experiments.
By using a parallelized version of our computer code the number of
atoms treated in the cluster can be easily increased to some hundreds. 
This is encouraging to extend the present calculations to larger
nanostructures (magnetic wires, dots, corrals etc.) 
currently being in the very focus of technological applications. 

\acknowledgements
This paper resulted from a collaboration partially funded
by the RTN network ``Computational Magnetoelectronics''
(Contract No. RTN1-1999-00145).
Financial support was also provided by the
Center for Computational Materials Science (Contract No. GZ 45.451),
the  Austrian Science Foundation (Contract No. W004),
and the Hungarian National Scientific Research Foundation
(OTKA T030240 and OTKA T029813).
We thank Prof. P.H. Dederichs for valuable discussions.

%%%%%%%%%%%%%%%%%%%%%%%%%%%%%%%%%%
%             TABLES              %
%%%%%%%%%%%%%%%%%%%%%%%%%%%%%%%%%%%
%     Fe Co Ni Moments            %
%%%%%%%%%%%%%%%%%%%%%%%%%%%%%%%%%%%
\begin{table}[h]{}
\caption{Calculated spin moments $(S_z)$ and orbital moments $(L_z)$, in units
 of \(\mu_B\),
for small clusters of Fe, Co and Ni on Ag(100) with
magnetization perpendicular to the surface ($z$).
{For each  position in a particular cluster 
(see Fig.~\protect{\ref{fig:clusters}})}, \(n_c\) refers to
the number of the neighboring magnetic (Fe, Co, Ni) atoms. 
}
%{\centering \begin{tabular}{|c|c|c||c|c||c|c||c|c|}
\label{tab:moments}
\vskip 0.5cm
{\centering 
\begin{tabular}{lcccccccc}
&&&\multicolumn{2}{c}{Fe}&\multicolumn{2}{c}{Co}&\multicolumn{2}{c}{Ni}\\
Cluster& position&$n_c$&
\( S_{z} \)& \( L_{z} \)& \( S_{z} \)& \( L_{z} \)& \( S_{z} \)& \( L_{z} \)\\
\hline
\hline
adatom &   & 0 & 3.39 & 0.88 & 2.10 & 1.19 & -- & -- \\
\\
dimer &1& 1& 3.31& 0.32& 2.09& 0.49&   --&   --\\
\\
trimer & 1 & 2 & 3.29 & 0.25 & 2.07 & 0.45 & 0.77 & 0.21 \\
       & 2 & 1 & 3.33 & 0.44 & 2.06 & 0.49 & 0.70 & 0.23 \\
\\
tetramer & 1 & 2 & 3.26 & 0.18 & 2.08 & 0.32 & 0.76 & 0.28 \\
\\
pentamer & 1 & 4 & 3.13 & 0.15 & 2.01 & 0.25 & 0.76 & 0.12 \\
         & 2 & 1 & 3.35 & 0.37 & 2.10 & 0.59 & 0.71 & 0.33 \\
         & 3 & 1 & 3.35 & 0.37 & 2.10 & 0.59 & 0.71 & 0.33 \\
\\
3$\times$3 cluster& 1 & 4 & 3.15 & 0.12 & 2.06 & 0.23 & 0.79 & 0.24\\
                  & 2 & 3 & 3.23 & 0.16 & 2.04 & 0.30 & 0.71 & 0.20\\
                  & 3 & 3 & 3.23 & 0.16 & 2.04 & 0.30 & 0.71 & 0.20\\
                  & 4 & 2 & 3.23 & 0.33 & 2.00 & 0.29 & 0.63 & 0.19 \\
\end{tabular}\par}
\end{table}
%
%%%%%%%%%%%%%%%%%%%%%%%%%%%%%%%%%%%
%           Fe Anisotropy         %
%%%%%%%%%%%%%%%%%%%%%%%%%%%%%%%%%%%
\begin{table}[h]{}
\caption{Calculated orbital moment anisotropies 
($\Delta L$), in units of \(\mu_B\), and contributions of the Fe atoms
to the MAE, $\Delta E$, in units of $meV$,
for small clusters of Fe on Ag(100). For each cluster, the total
MAE per Fe atom of the cluster including the neighborhood
is also given in parentheses.
}
\label{tab:feani}
\vskip 0.5cm
{\centering \begin{tabular}{lcccccc}
Cluster&
position&$n_c$&
\(\Delta L_{x-z} \)& \( \Delta E_{x-z} \) &
\( \Delta L_{y-z} \)& \( \Delta E_{y-z} \)
\\
\hline
\hline
adatom &   & 0 & -0.37 &  5.07  & -0.37 &  5.07  \\
       &   &   &       & (5.61) &       & (5.61) \\
\\
dimer  & 1 & 1 & -0.12 &  2.14  & -0.11 &  1.66  \\
       &   &   &       & (2.30) &       & (1.83) \\
\\
trimer & 1 & 2 & -0.12 &  1.93  & -0.08 &  0.93  \\
       & 2 & 1 & -0.16 &  2.83  & -0.15 &  2.39  \\
       &   &   &       & (2.72) &       & (2.13) \\
\\
tetramer & 1 & 2 & -0.02 &  0.50  & -0.02 &  0.50 \\
         &   &   &       & (0.54) &       & (0.54) \\
\\
pentamer  & 1 & 4 & -0.03 &  0.49  & -0.03 &  0.49 \\
          & 2 & 1 & -0.03 &  0.92  & -0.08 &  0.85 \\
          & 3 & 1 & -0.08 &  0.85  & -0.03 &  0.92 \\
          &   &   &       & (0.90) &       & (0.90)\\
\\
 3$\times$3 cluster& 1 & 4 &  0.00 &  0.23  &  0.00 &  0.23  \\
                   & 2 & 3 & -0.02 &  0.43  & -0.01 &  0.84  \\
                   & 3 & 3 & -0.01 &  0.84  & -0.02 &  0.43  \\
                   & 4 & 2 & -0.13 &  1.86  & -0.13 &  1.86  \\
                   &   &   &       & (1.20) &       & (1.20) \\
\end{tabular}\par}
\end{table}

%%%%%%%%%%%%%%%%%%%%%%%%%%%%%%%%%%%
%           Co Anisotropy         %
%%%%%%%%%%%%%%%%%%%%%%%%%%%%%%%%%%%
\begin{table}[h]{}
\caption{As in Table \protect{\ref{tab:feani}} for Co clusters.}
\vskip 0.5cm
{\centering \begin{tabular}{lcccccc}
Cluster& position&$n_c$&
\(\Delta L_{x-z} \)& \( \Delta E_{x-z} \) &
\( \Delta L_{y-z} \)& \( \Delta E_{y-z} \)
\\
\hline
\hline
adatom &   & 0 & -0.26 &  4.20  & -0.26 &  4.20  \\
       &   &   &       & (4.36) &       & (4.36) \\
\\
dimer  & 1 & 1 &  0.15 & -3.50  & -0.01 &  0.67  \\
       &   &   &       &(-3.49) &       & (0.76) \\
\\
trimer & 1 & 2 &  0.40 & -9.06  & -0.02 & -0.11  \\
       & 2 & 1 & 0.34  & -6.29  &  0.05 & -0.04  \\
       &   &   &       &(-7.44) &       & (-0.01)\\
\\
tetramer & 1 & 2 & 0.15 & -2.29 & 0.15 & -2.29  \\
         &   &   &      &(-2.37)&      &(-2.37) \\
\\
pentamer & 1 & 4 & 0.12  & -2.46  &  0.12 & -2.46 \\
         & 2 & 1 & 0.21  & -4.16  & -0.01 & -0.03 \\
         & 3 & 1 & -0.01 & -0.03  &  0.21 & -4.16 \\
         &   &   &       &(-2.22) &       &(-2.22)\\
\\
3$\times$3 cluster& 1 & 4 &  0.13 & -1.86 & 0.13 & -1.86 \\
                  & 2 & 3 &  0.10 & -1.56 & 0.18 & -2.96 \\
                  & 3 & 3 &  0.18 & -2.96 & 0.10 & -1.56 \\
                  & 4 & 2 &  0.16 & -2.60 & 0.16 & -2.60 \\
                  &   &   &       &(-2.45)&      &(-2.45)\\
\end{tabular}\par}
\label{tab:coani}
\end{table}

%%%%%%%%%%%%%%%%%%%%%%%%%%%%%%%%%%%
%           Ni Anisotropy         %
%%%%%%%%%%%%%%%%%%%%%%%%%%%%%%%%%%%
\begin{table}[h]{}
\caption{As in Table \protect{\ref{tab:feani}} for Ni clusters.}
\vskip 0.5cm
{\centering \begin{tabular}{lcccccc}
Cluster&
position&$n_c$&
\(\Delta L_{x-z} \)& \( \Delta E_{x-z} \) &
\( \Delta L_{y-z} \)& \( \Delta E_{y-z} \)
\\
\hline
\hline
trimer & 1 & 2 & 0.19 & -6.12  & 0.18 & -1.38  \\
       & 2 & 1 & 0.11 & -3.72  & 0.08 & -1.00  \\
       &   &   &      &(-4.63) &      &(-1.13) \\
\\
tetramer & 1 & 2 & -0.05 & 0.07 & -0.05  & 0.07  \\
         &   &   &       &(0.10)&        &(0.10) \\
\\
pentamer & 1 & 4 & 0.15 & -2.26 & 0.15 & -2.26 \\
         & 2 & 1 & 0.01 & -1.64 & 0.05 & -0.69 \\
         & 3 & 1 & 0.05 & -0.69 & 0.01 & -1.64 \\
         &   &   &      &(-1.41)&      &(-1.41)\\
\\
3$\times$3 cluster & 1 & 4 & -0.06 & -0.02 & -0.06 &  -0.02  \\
                   & 2 & 3 &  0.02 & -0.75 &  0.06 &  -2.00  \\
                   & 3 & 3 &  0.06 & -2.00 &  0.02 &  -0.75  \\
                   & 4 & 2 &  0.05 & -1.21 &  0.05 &  -1.21  \\
                   &   &   &       &(-1.17)&        & (-1.17)\\
\end{tabular}\par}
\label{tab:niani}
\end{table}

%%%%%%%%%%%%%%%%%%%%%%%%%%%%%%%%%%%
%           FIGURES               %
%%%%%%%%%%%%%%%%%%%%%%%%%%%%%%%%%%%
%%%%%%%%%%%%%%%%%%%%%%%%%%%%%%%%%%%
%          CLUSTERS                %
%%%%%%%%%%%%%%%%%%%%%%%%%%%%%%%%%%%
{\centering
\begin{figure}[h]{}
%%%  TWO-COLUMN FORMAT and EPSF %%%%%%%
\epsfxsize=7cm \centerline{\epsffile{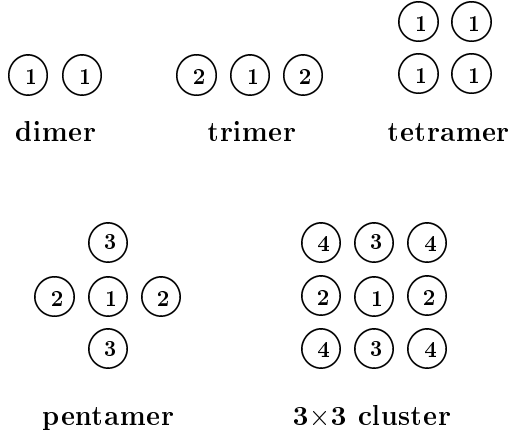}} \vspace{0.3cm}
%%%  TWO-COLUMN FORMAT and EPSF %%%%%%%
\caption{Sketch of the planar clusters considered.
For an orientation of the  magnetization along the $x$ or $y$ axis, 
the equivalent atoms in a cluster are labelled by the same number.}
\label{fig:clusters}
\end{figure}
}
%
%%%%%%%%%%%%%%%%%%%%%%%%%%%%%%%%%%%
%           SHELL-TEST             %
%%%%%%%%%%%%%%%%%%%%%%%%%%%%%%%%%%%
{\centering
\begin{figure}[h]{}
%%%  TWO-COLUMN FORMAT and EPSF %%%%%%%
\epsfxsize=7cm \centerline{\epsffile{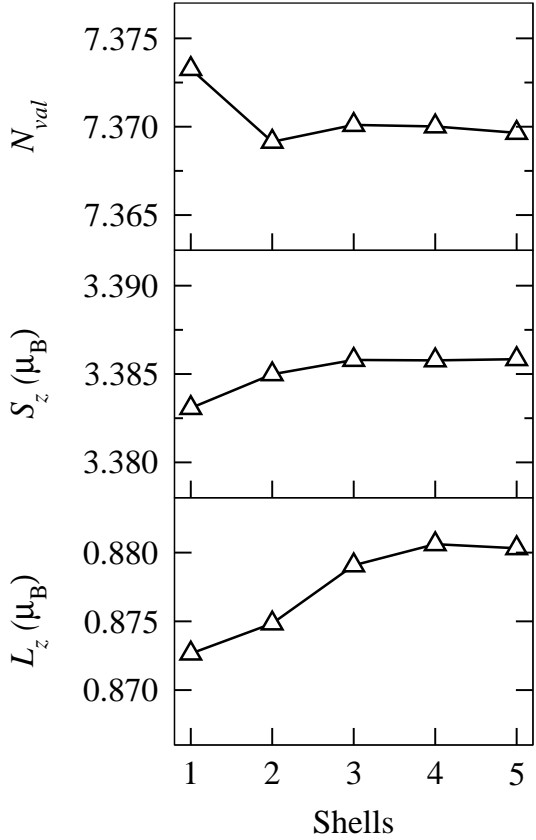}} \vspace{0.3cm}
%%%  TWO-COLUMN FORMAT and EPSF %%%%%%%
\caption{Calculated number of valence electrons ($N_{val}$), spin 
moment ($S_z$) and orbital moment ($L_z$) of a single Fe adatom on a
Ag(100) surface as a function of the number 
of the self-consistently treated atomic shells around the Fe atom.}
\label{fig:shells}
\end{figure}
}
%
%
%%%%%%%%%%%%%%%%%%%%%%%%%%%%%%%%%%%
%     2 Fe, Co MOMENTS            %
%%%%%%%%%%%%%%%%%%%%%%%%%%%%%%%%%%%
{\centering
\begin{figure}[h]{}
%%%  TWO-COLUMN FORMAT and EPSF %%%%%%%
\epsfxsize=7cm \centerline{\epsffile{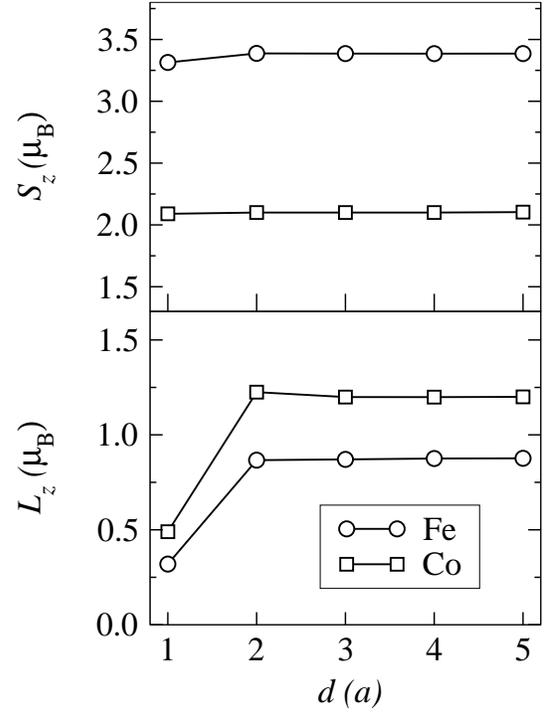}} \vspace{0.3cm}
%%%  TWO-COLUMN FORMAT and EPSF %%%%%%%
\caption{Calculated spin and orbital moments of two adatoms of Fe or Co
on Ag(100) as a function of their distance $d$ measured
in units of the 2D lattice constant $a$.}
\label{fig:moments2feco}
\end{figure}
}
%
%
%%%%%%%%%%%%%%%%%%%%%%%%%%%%%%%%%%%
%  EXCHANGE-COUPLING ENERGY       %
%%%%%%%%%%%%%%%%%%%%%%%%%%%%%%%%%%%
{\centering
\begin{figure}[h]{}
%%%  TWO-COLUMN FORMAT and EPSF %%%%%%%
\epsfxsize=7cm \centerline{\epsffile{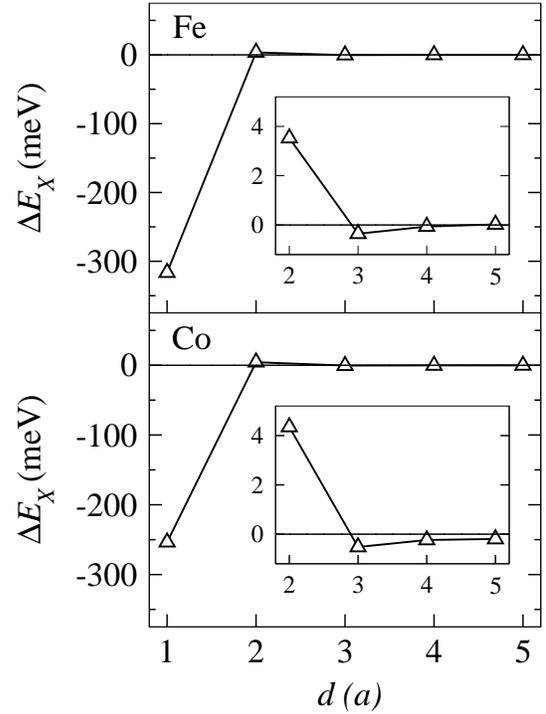}} \vspace{0.3cm}
%%%  TWO-COLUMN FORMAT and EPSF %%%%%%%
\caption{Calculated exchange coupling energy, $\Delta E_{X}$, between two
adatoms of Fe or Co on Ag(100) as a function of the distance $d$
measured in units of the 2D lattice constant $a$.
The insets show the range $2a \le d \le 5a$ on a blown up scale.
}
\label{fig:xc}
\end{figure}
}

\end{document}